\documentclass{pasj00}
\begin{document}
\SetRunningHead{K. Sadakane et al.}{Abundances in the Secondary Star of V4641 Sgr}
\Received{2006 January 12}
\Accepted{2006 April 12}

\title{Chemical Abundances in the Secondary Star of the Black Hole Binary V4641 Sagittarii
  (SAX J1819.3--2525)
\thanks{Based on data collected at the Subaru Telescope,
                  which is operated by the National Astronomical Observatory
of Japan.}}

\author{Kozo \textsc{Sadakane},\altaffilmark{1}
        Akira \textsc{Arai},\altaffilmark{1}
        Wako \textsc{Aoki},\altaffilmark{2}
        Nobuo \textsc{Arimoto},\altaffilmark{2}
        Masahide \textsc{Takada-Hidai},\altaffilmark{3}
        Takashi \textsc{Ohnishi},\altaffilmark{4} \\
        Akito \textsc{Tajitsu},\altaffilmark{5}
        Timothy C. \textsc{Beers},\altaffilmark{6}
        Nobuyuki \textsc{Iwamoto},\altaffilmark{7}
        Nozomu \textsc{Tominaga,}\altaffilmark{8}
        Hideyuki \textsc{Umeda},\altaffilmark{8} \\
        Keiichi \textsc{Maeda},\altaffilmark{9}
        and Ken'ichi \textsc{Nomoto}\altaffilmark{8}}

\altaffiltext{1}{Astronomical Institute, Osaka Kyoiku University, Asahigaoka,  Kashiwara, Osaka
         582-8582}
\altaffiltext{2}{National Astronomical Observatory, 2-21-1 Osawa,
         Mitaka, Tokyo 181-8588}
\altaffiltext{3}{Liberal Arts Education Center, Tokai University, 1117 Kitakaname, Hiratsuka, 
         Kanagawa 259-1292}
\altaffiltext{4}{Nagoya Science Museum, 2-17-1, Sakae, Naka-ku, Nagoya, Aichi 460-0008}
\altaffiltext{5}{Subaru Telescope, National Astronomical Observatory of Japan,
           650 North A'ohoku Place, \\
           Hilo, HI 96720, USA}
\altaffiltext{6}{Department of Physics and Astronomy and Joint Institute for Nuclear
           Astrophysics (JINA), \\
           Michigan State University, East Lansing, MI 48824, USA}
\altaffiltext{7}{Nuclear Data Center, Japan Atomic Energy Agency, 
            Ibaraki 319-1195}
\altaffiltext{8}{Department of Astronomy, School of Science, University of Tokyo, 
           Bunkyo-ku, Tokyo 113-0033}
\altaffiltext{9}{Department of Earth Science and Astronomy,
           College of Arts and Sciences, 
           University of Tokyo, Tokyo 153-8902}
\email{sadakane@cc.osaka-kyoiku.ac.jp}


\maketitle

\begin{abstract}

We report on detailed spectroscopic studies performed for the secondary star in the
black hole binary (micro-quasar) V4641 Sgr in order to examine its surface
chemical composition and to see if its surface shows any signature of pollution
by ejecta from a supernova explosion. High-resolution spectra of V4641 Sgr
observed in the quiescent state in the blue-visual region are compared with
those of the two bright well-studied B9 stars (14 Cyg and $\nu$ Cap) observed
with the same instrument. The effective temperature of V4641 Sgr (10500 $\pm$
200 K) is estimated from the strengths of He~{\sc i} lines, while its rotational
velocity, $\it v$ sin $\it i$ (95 $\pm$ 10 km s${}^{-1}$), is estimated from the
profile of the Mg~{\sc ii} line at 4481 \AA. We obtain abundances of 10 elements
and find definite over--abundances of N (by 0.8 dex or more) and Na (by 0.8 dex)
in V4641 Sgr. From line-by-line comparisons of eight other elements (C, O, Mg,
Al, Si, Ti, Cr, and Fe) between V4641 Sgr and the two reference stars, we
conclude that there is no apparent difference in the abundances of these elements
between V4641 Sgr and the two normal late B-type stars, which have been reported to
have solar abundances. An evolutionary model of a massive close binary system
has been constructed to explain the abundances observed in V4641 Sgr.
The model suggests that the progenitor of the black hole forming supernova
was as massive as $\sim 35 M_\odot$ on the main-sequence and, after becoming a
$\sim 10 M_\odot$ He star, underwent "dark" explosion which ejected only N and
Na-rich outer layer of the He star without radioactive $^{56}$Ni.    

\end{abstract}

\section {Introduction}

The object V4641 Sgr is an X-ray binary with an orbital period of 2.817 d,
containing a primary (a black hole) of $\sim$ 9.6 \Mo~ and a secondary star of 5
-- 8 \Mo \citep{orosz01}. Extensive photometric observations were carried out by
\citet{goran03}, who reported a photometric period of 2.81728 d, and $\it V$ and
$\it R$ bands light curves which were dominated by ellipsoidal variations from
the secondary. They demonstrated that the surface temperature of the secondary
is non--uniform from the observed color variation. They observed a depression in
the red wing of the H$\alpha$ line just before the black hole's inferior
conjunction, and interpreted this as being  due to absorption by the rarefied gas disk
around the black hole.

The source exhibited two rapid and bright X-ray flares around 1999 September 15
\citep{smith99}. Just prior to the giant X-ray flare, the source showed an
increase in the variability \citep{kato99}. About six days before the X-ray
flare, the source exhibited a $\sim$ 1 mag modulation with a period of $\sim$ 2.5
d. After reaching the peak brightness (8.8 mag in $\it V$), it decayed
rapidly to its mean quiescent level within two days. Since then, the source
exhibited several outbursts with very rapid fluctuations in the optical band
(Uemura et al. 2002a, 2002b, 2004a, 2004b, 2005).

\citet{orosz01} carried out extensive spectroscopic observations of V4641 Sgr in
1999 and obtained orbital parameters of the system. The spectral type of the
secondary star was assigned to be B9 III. They determined the rotational
velocity, $\it v$ sin $\it i$, to be 123 $\pm$ 4 km s${}^{-1}$, and performed
spectroscopic analyses of the secondary star using moderate-resolution spectra
to estimate the abundances of several light elements. They found over--abundances of
several elements including N, O, Mg, Ca, and Ti, with a solar abundance of
Si. Although they note that better data (e.g., high-resolution spectra) or
better models are needed before establishing the abundance anomalies, their
results on (possible) over--abundances of light elements are very interesting
because chemical abundances of the secondary star are expected to provide direct
information on the products of nucleosynthesis from supernova explosions of
massive stars. Information on abundances is also expected to constrain many
phisical parameters that are involved in supernova explosion models. They
include the mass cut, the amount of fall-back matter, possible mixing, and
explosion energies and geometries.

\citet{isra99} reported over--abundances of O,  Mg, Si, S, and Ti, but a
solar abundance of Fe, in the eclipsing low--mass X-ray binary GRO J1655-40
(Nova Scorpii 1994). Based on this observation and using a variety of supernova
models, \citet{podsi} showed that the best fits can be obtained for He star
masses of 10 -- 16 \Mo, where spherical hypernova models are generally favored
over standard supernova ones. Over--abundances of Al, Ti, and Ni are reported in
the black hole binary A0620--00 by \citet{gonz04}. They discussed a possible
scenario of pollution from a supernova. \citet{gonz05} found enhanced abundances
of Ti, Fe, and Ni in the neutron star binary Cen X-4, and showed that these
apparent anomalies can be explained if the secondary star captured a significant
amount of matter ejected from a spherically symmetric supernova explosion of a 4
\Mo~ He core progenitor.

In order to clarify the possible abundance anomalies reported in V4641 Sgr, we
 carried out high-resolution spectroscopic observations using the Subaru
telescope and performed comparative abundance analyses using two well-studied
reference stars that are reported to have solar abundances.
 
\section {Observational Data}

Spectroscopic observations of V4641 Sgr were carried out with the Subaru
Telescope using the High Dispersion Spectrograph (HDS) on 2005 May 21 and June
19 (UT). Our observations were made during its quiescent state, though the
object has been reported to exhibit an outburst on 2005 June 24.371 (UT) (M.
Uemura, private communication). Data for two reference stars [14 Cyg = HD 185872
(B9 III) and $\nu$ Cap = HD193432 (B9.5 V)] were obtained on 2005 May 20 and
June 19, respectively, using the same instrumental setup. Technical details
and the performance of the spectrograph are described in \citet{nogu02}. We used
a slit width of \timeform{1".0} (0.5 mm) and a 2x2 binning mode, which enabled
us to achieve a nominal spectral resolving power of about $R = 45000$ with
a 3.5 pixel sampling. Our observations covered the wavelength region from 4050 \AA~ to 6760
\AA~with a gap between 5340 -- 5450 \AA. The exposure times for V4641 Sgr were 4400
sec and 5400 sec on May 21 and June 19, respectively. For flat-fielding of the
CCD data, we obtained Halogen lamp exposures (flat images) with the same setup
as that for the object frames.

The reduction of two-dimensional echelle spectral data was performed using the
IRAF software package in a standard manner. Spectral data extracted from
multiple object images of V4641 Sgr were averaged in order to improve the
signal-to-noise (S/N) ratio. The wavelength calibration was performed using the
Th-Ar comparison spectra obtained during the observations. The measured FWHM of
the weak Th lines was 0.15 \AA~ at 6000 \AA, and the resulting resolving power was
around $R = 40000$.

When the observed star shows broadened line profiles due to its high rotational
velocity, the process of continuum fitting to the extracted raw spectral data
has to be carried out carefully. A very shallow and wide spectral feature might
be mistakenly interpreted as the continuum level when a high-order polynomial
function is employed in the process. The task of continuum fitting the spectral
data for V4641 Sgr was carried out independently by two of us (K. S. and M.
T.-H.) using two different approaches. K. S. saved the fitted functions of each
echelle order obtained for the spectra of the sharp-lined stars 14 Cyg and $\nu$
Cap, and used them to divide  the raw spectral data of the corresponding
echelle orders of V4641 Sgr. Fitted functions of the continuum for 14 Cyg and
$\nu$ Cap were then applied to the V4641 Sgr observations obtained on May 21 and
June 20, respectively. M. T.-H. carefully selected line-free windows in each
echelle order for spectra of the rapidly rotating spectrophotometric standard
star Feige 15 \citep{hidai02}, and used them as reference to determine line-free
windows in echelle spectra of V4641 Sgr. Since
Feige 15 is classified as a spectral type of A0 in the SIMBAD database (which is
operated at CDS, Strasbourg, France), its effective temperature can be regarded
as being very close to that of V4641 Sgr, with a spectral type of B9 III,
suggesting a spectroscopic similarity between the two stars. We found good
agreement between the results obtained from the two methods, even for very weak
features, and concluded that we had correctly obtained normalized spectral data
to be used for the abundance analysis.

We compared the spectra of V4641 Sgr obtained on two nights [May 21 and June 19, at
photometric phases 0.38 and 0.62, respectively, which are calculated using data
given in \citet{goran03}] and found no detectable spectral variation. Thus, the data
observed on the two nights were averaged in order to obtain the final spectrum, after
correcting for the apparent doppler shifts due to the orbital motion. The S/N
ratios of the resulting spectrum of V4641 Sgr were measured at several continuum
windows near 5000 \AA~ and 6000 \AA . The averaged S/N ratio (per pixel) were
around 180 near 5000 \AA~ and 210 near 6000 \AA. For the two reference stars,
a much higher S/N ratio (around 400) was achieved at 6000 \AA .

\section {Spectral Analysis}

An abundance analysis of V4641 Sgr has been carried out relative to the two
reference stars 14 Cyg and $\nu$ Cap. Table 1 lists the absorption lines
used in the following analyses. Log $\it gf$ values are taken from the VALD
database \citep{kupka}; atmospheric parameters ({\it $T_{\rm eff}$} and log {\it
g}) for these two stars were taken from \citet{adel02}. They are ({\it $T_{\rm
eff}$} and log {\it g}) = (10750 K, 3.5) and (10250 K, 4.0) for 14 Cyg and $\nu$
Cap, respectively. Abundance analyses of these two stars have been published in
\citet{adel99} (14 Cyg) and in \citet{adel91}($\nu$ Cap); solar abundances are
reported in these two stars.

We estimated the effective temperature of V4641 Sgr by comparing the strengths
of the two He~{\sc i} lines at 4471 \AA~ and 5876 \AA~ in V4641 Sgr and in the
two reference stars, as shown in figure 1. In the figure, the observed spectra
of these three stars are compared with simulated ones. Spectral simulations were
performed using line-blanketed model atmospheres interpolated from
\citet{kuru93} and assuming the solar abundances (except for figures 2 and 8, in
which an enhanced abundance of Mg by 0.20 dex is assumed). In the spectral
simulations, atomic data, except for the major features listed in table 1, are
taken from \citet{kubell95}. Equivalent widths of the two He~{\sc i} lines were
measured by directly integrating their profiles in the three stars. We found
that equivalent widths of both of the He~{\sc i} lines in V4641 Sgr are just
around the average of the two reference stars and conclude that the {\it $T_{\rm
eff}$} of V4641 Sgr is near the average of the two stars (10500 $\pm$ 200 K).
This is in agreement with that obtained in \citet{orosz01}.

We adopt the surface gravity (log {\it g} = 3.5) of V4641 Sgr given in
\citet{orosz01}, who obtained this value from the measured widths of the Balmer
lines. We use a microturbulent velocity $\xi$${}_{\rm t}$ = 1.6 km s${}^{-1}$ in
$\nu$ Cap \citep{dwo00}, and assume $\xi$${}_{\rm t}$ = 2.0 km s${}^{-1}$ in 14
Cyg. It is difficult to determine the microturbulent velocity in V4641 Sgr
because we cannot measure the weak metallic lines, which are in any case
unaffected by a change in $\xi$${}_{\rm t}$, in this star. The microturbulent
velocity was guessed to be around 2.5 $\pm$ 0.5 km s${}^{-1}$ by simulating the
spectral region between 4505 and 4535 \AA~, where moderately strong (80 - 150
m\AA) Fe~{\sc ii} lines can be found, to obtain the best fitting, assuming an
appropriate rotational velocity ($\it v$ sin $\it i$, described in the next
section).

\section{Results}

We estimate the rotational velocity,  $\it v$ sin $\it i$, of V4641 Sgr
from the observed profile of the Mg~{\sc ii} line at 4481 \AA~, as shown in
figure 2. First, we tried to reproduce the profiles of the line in 14 Cyg and $\nu$ Cap
by adopting the mean values of published data of $\it v$ sin $\it i$. They are
30 and 25 km s${}^{-1}$ for 14 Cyg (\cite{abt95b}; \cite{royer02}) and $\nu$ Cap
(\cite{abt95a}; \cite{dwo00}; \cite{fekel03}; \cite{royer02}), respectively. We
find that acceptable reproductions of the observed line profile can be obtained
using adopted values of $\it v$ sin $\it i$ for the two stars, as shown in
figure 8. Next, we searched for a suitable value of $\it v$ sin $\it i$ in V4641
Sgr by exploring values between 60 and 150 km s${}^{-1}$; we found that the best
fit is achieved at 95 $\pm$ 10 km s${}^{-1}$. If we use $\it v$ sin $\it i$ =
123 km s${}^{-1}$, as obtained in \citet{orosz01}, the simulated profile becomes
too shallow and too wide (figure 2). We conclude that their result of $\it v$
sin $\it i$ is too large, most probably resulting from the low resolution of
their spectral data.

We tried to estimate the abundance of C in V4641 Sgr using the C~{\sc ii} line
at 4267 \AA~ (figure 3). The line consists of two major components at 4267.001
\AA~ and at 4267.261 \AA~ (table 2). The C~{\sc ii} feature in $\nu$ Cap can be
reproduced with a solar abundance of C, while the observed feature in 14 Cyg
appears to be too weak. This may suggest either a slight under--abundance of C
or an error in the adopted atmospheric parameters for 14 Cyg. We find that the
broard feature at 4267 \AA~ in V4641 Sgr can be reproduced by assuming a solar
abundance of C.

We find a shallow and broard absorption feature at 6483 \AA~ in V4641 Sgr in
both the May 20 and June 19 data. Comparisons with the two reference stars
(figure 4) shows that the feature in V4641 Sgr appears to be too deep. The
feature coincides with the positions of the five components of N~{\sc i} lines
(Multiplet No. 21). We simulated spectra of both reference stars, assuming solar
abundances, and find that observations can be reasonably reproduced when
excluding the sharp components due to atmospheric absorption. The only
observable feature is the Fe~{\sc ii} line at 6482.204 \AA. On the other hand,
the observed feature of V4641 Sgr at 6483 \AA~ cannot be reproduced when we
assume solar abundances of N and Fe. A reasonable reproduction can be obtained
when a significant over--abundance of N (by around 1.0 dex) is assumed, as shown
in figure 4. Allowing for a slight uncertainty in the continuum level because of
the low S/N ratio, we conclude that N is indeed over--abundant in V4641 Sgr by
at least 0.8 dex. This conclusion is in accordance with the result noted in
\citet{orosz01}, who obtained an over--abundance of N by 1.0 dex from an
analysis of the same spectral feature.
   
Analyses of weak O~{\sc i} lines near 6455 \AA~ are shown in figure 5. We
conclude that the O~{\sc i} feature can be reasonably fit by using a solar O
abundance. This is in contrast with the result given in \citet{orosz01}, who
obtained an over--abundance of O (by a factor of three), when analysing the same
spectral feature.

Next, we analyse the Na~{\sc i} D lines. The D lines are usually contaminated by
interstellar absorption superposed on the stellar components, as shown for 14
Cyg and $\nu$ Cap (figure 6). Fortunately, spectral lines of V4641 Sgr observed
on July 19 are significantly redshifted (by +210 km s${}^{-1}$), and we can
analyse both of the D1 and D2 components, which are unaffected by interstellar
absorptions. The broad D lines in V4641 Sgr are too strong to be accounted for
by the solar Na abundance. As is shown in figure 6, we need to enhance the
abundance of Na by 0.7 dex in order to fit the D2 line. Further enhancement (by
0.2 dex) of the Na abundance is needed to fit the D1 line (although the D1 line
is heavily affected by atmospheric absorption). Upon 
averaging the results obtained from
the D1 and D2 lines, we concluded that Na is over--abundant by at least 0.8 dex
in V4641 Sgr.

The abundances of six elements (Mg, Si, Al, Ti, Cr, and Fe) in V4641 Sgr were
estimated by comparing the major absorption features of each element with those
seen in the two reference stars. figures 7 to 11 display the results. In these
figures, the solid lines show simulated spectra for each star, assuming solar
abundances (except for figure 8, where an enhanced abundance of Mg is assumed),
taken from \citet{gre98}. Neutral and singly ionized lines of Mg are shown in
figures 7 and 8, respectively. We find that the Mg~{\sc i} triplet lines can be
reproduced with a solar abundance of Mg, while a slight enhancement (0.20 dex)
of Mg is suggested from the observed profile of the Mg~{\sc ii} line at 4481
\AA. On the other hand, the Mg~{\sc ii} feature at 4390.5 \AA~ in V4641 Sgr can
be reproduced by assuming a solar Mg abundance. Thus, we conclude that the
abundance of Mg in V4641 Sgr coincides with that of the Sun, which differs from
the results reported by \citet{orosz01}, who concluded that this star exhibits
an enhanced (by a factor of seven) abundance of Mg.

We find that an Al~{\sc ii} line at 4663 05 \AA~ is clearly present in both 
of the two reference stars, and in V4641 Sgr (figure 9). 
Our spectral analyses of this line suggests
that Al may by slightly over--abundant (by +0.2 dex) in V4641 Sgr, although the
relatively poor S/N ratio in the blue region does not allow for a reliable analysis of
the line. A pair of Si~{\sc ii} lines near 5050 \AA~ are compared in figure 10.
We conclude that the abundance of Si in V4641 Sgr is close to the solar
abundance. In figure 11, we compare Ti~{\sc ii}, Cr~{\sc ii}, and Fe~{\sc ii}~
lines, where all of the absorption features can be reproduced by assuming the same
(solar) abundances in the three stars. We analyse several other strong Ti~{\sc ii},
Cr~{\sc ii}, and Fe~{\sc ii}~ lines (noted in table 1) and found that all of these
features in V4641 Sgr can be explained using solar abundances of Ti, Cr, and
Fe. Our result for Ti is again in contrast with the result given in
\citet{orosz01}, who obtained an over--abundance of Ti (by a factor of 10). They
used four Ti~{\sc ii} lines (at 5129.2 \AA, 5185.9 \AA, 5188.7 \AA, and at
5226.5 \AA) to obtain the Ti abundance, and pointed out a high abundance of Ti
from the two lines (at 5129.2 \AA~ and 5226.5 \AA) using data of spectral
resolution of about 4 \AA. We examined all of these lines on our high-resolution data and
find that all these lines are very weak in the reference stars, and also in V4641
Sgr, when compared to those Ti~{\sc ii} lines listed in table 1. We concluded that
the four lines used in \citet{orosz01} are inadequate to be used for the
abundance analysis of Ti.

Our final derived abundances for 10 observed elements (11 ions) are summarized
in table 2, together with their expected errors. We estimate uncertainties in the abundances of
each ion introduced by errors in the adopted parameters: 200 K in {\it $T_{\rm
eff}$}, 0.5 in log $\it g$, and 0.5 km s${}^{-1}$ in $\xi$${}_{\rm t}$.  When
these errors are combined, we conclude that the our abundance analysis results
are reliable within 0.25 dex (table 2).
We examined the effect of a difference in spectral resolution on the resulting
abundances by a simple test. The original data of both 14 Cyg and V4641 Sgr were
degraded to around $R = 8000$, the highest resolution used in \citet{orosz01}, by
convolving with an appropriate Gaussian function. We then repeated abundance analyses using
the degraded data for several spectral features such as the Mg~{\sc ii} line at 4481 \AA~,
the Mg~{\sc i} triplet lines, and the three  Ti~{\sc ii} lines listed in table 1.
Fairly good agreements (within 0.05 dex) were found for 14 Cyg from both high and
low resolution data. On the other hand,
differences as large as 0.15 dex were found between abundance results obtained from
 weak and noisy spectral features in the case of  V4641 Sgr.
We infer that  these differences are mainly resulted from the relatively poor
 SN ratio in the V4641 Sgr data, but not from the difference in the spectral resolution.
When the limited S/N ratio of our
observation and the high rotational velocity of V4641 Sgr are taken into account, the
expected error in the abundances should be increased to around 0.3 dex.

\section{Discussion}

We obtained abundances of 10 elements in V4641 Sgr, and found definite
over--abundances of only two elements (N and Na). The abundances of the eight other
elements in V4641 Sgr have been shown to be the same as those in the two reference
stars (solar abundances), except for a possible enhancement of Mg suggested from
the Mg~{\sc ii}~line at 4481 \AA~ and that of Al. However, when averaged with
the result obtained from the Mg~{\sc i} triplet lines and the Mg~{\sc ii} line
at 4390.5 \AA, the abundance of Mg is coincident with that in the reference
stars within the expected error ([Mg/H] = +0.10 $\pm$ 0.30). The above
conclusions are in contrast to the results noted in \citet{orosz01} except for
N, where they concluded over--abundances of N (1.0 dex), O
(0.48 dex), Mg (0.85 dex), and Ti (1.0 dex) in V4641 Sgr when compared to the Sun.
We suggest that the primary reason for obtaining discordant abundances for
O, Mg, and Ti is the difference in the spectral resolution of the data.
\citet{orosz01} used a much lower resolving power ($\it R$ ranging from 1200 to 7700)
than obtained in the present study ($\it R$ $\sim$ 40000). 

Our results for the abundances of the light elements in the secondary star of V4641
Sgr [definite over--abundances of N and Na, normal (solar) abundances of O, and
the $\alpha$-elements Mg, Si, and Ti] are unique when compared with the
results obtained for other X-ray binaries. Abundances obtained in four secondary
stars in X-ray binaries are compared in figure 12 [V4641 Sgr (this study), GRO
J1655-40 \citep{isra99}, A0620-00 \citep{gonz04}, and Cen X4 \citep{gonz05}]. We
note that all four stars show distinct abundance patterns. The $\alpha$-elements
(O, Mg, Si, S, and Ti) are definitely over--abundant in GRO
J1655-40, while they appear to be normal in V4641 Sgr. Fe is over--abundant only
in Cen X-4. N and Na are over--abundant only in V4641 Sgr. The difference in the
abundances of O between V4641 Sgr and GRO J1655-40 is impressive.
The observed abundance pattern in V4641 Sgr (enhanced N and Na, and
normal $\alpha$-elements) seems to be different from those of the
usual supernova models that predict the enhancement of
$\alpha$-elements (\cite{podsi}, \cite{gonz04}, and \cite{gonz05}).
However, these variations of abundance patters can be explained with
the variations of the abundances of supernova explosions that are
associated with black hole formation \citep{umeda03}.

In order to explain the abundances in V4641 Sgr quantitatively, we
 calculated the evolution of the star with the initial mass of 40
$M_\odot$ and the solar metallicity from the main-sequence to collapse
as in \citet{umeda05} and constructed the following evolutionary
models for V4641 Sgr.
In the close binary system, this 40 $M_\odot$ primary star underwent a
common envelope phase and lost most of its H-rich envelope until it
became a He star of mass 15.14 $M_\odot$.  The system also lost its
angular momentum and became compact with the orbital period as short
as that observed.
Figure 13 shows the abundance distribution near the surface of
the He star at the onset of collapse.  In the He-rich layer, $^{14}$N
and $^{23}$Na were enhanced by the CNO-cycle and Ne-Na cycle (proton
captures on $^{21}$Ne and $^{22}$Ne) during H-burning.  In the deeper
He layer, the $^{14}$N abundance was decreased by successive
$\alpha$-capture to produce $^{22}$Ne during weak He shell burning.

The 15 $M_\odot$ He star is massive enough to eventually formed a
black hole (BH).  We assume that the collapse induced a relatively
weak explosion.  Generally, an explosion with a smaller energy leads
to a larger amount of fall back materials and thus a smaller amount
of ejecta (e.g., \cite{iwamoto05}).  In order to reproduce the
observations of V4641 Sgr, we assume that the kinetic energy of
explosion was as small as $E=6\times 10^{49}$ ergs.  In such a weak
explosion, only 0.5 $M_\odot$ materials above $M_r=14.66M_\odot$ were
ejected.  The abundance distribution in the He layer in figure 13
does not change in the explosion.
A part of the ejected materials must be captured by the secondary
star.  The captured (accreted) materials were then mixed with the
materials of solar metallicity in the atmosphere of the secondary
star.  Since the ejecta is relatively N- and Na-rich without any
enhancement of $\alpha$-elements, this could explain the observed
abundance pattern of V4641 Sgr.  If the accreted material is mixed
with 40-times larger amount of secondary star materials, the final
abundance pattern would be consistent with the observed abundance pattern of
the secondary star (figure 14).  Here, most of the heavy elements
above Na originated from the materials of V1641 Sgr.  Such a partial
mixing (e.g., slow rotational mixing) may be realized because the
surface temperature of V4641 Sgr is
too high for deep convective mixing to occur.

We should note that an alternative scenario is possible.  If the
stellar wind of the He star of $\sim$ 15 $M_\odot$ blows at a high
enough rate, a part of the N- and Na-rich materials in the He layer
would have been blown off and captured by the secondary star.  If the energy of
supernova explosion was even smaller than $\sim 6 \times 10^{49}$, no
mass ejection occurred.  These processes could lead to the observed
abundance pattern of V4641 Sgr.
The above 40 $M_\odot$ model formed a BH of 14 $M_\odot$.  The
initially 30 $M_\odot$ model produces a similar abundance pattern by
forming a 7.2 $M_\odot$ BH.  Since the observed BH mass of V4641 Sgr
is $\sim 9.6 M_\odot$, the progenitor of the BH in V4641 is likely a
$\sim 35 M_\odot$ star.
It is highly uncertain in the current supernova models under what
condition the BH formation can induce a supernova explosion and how
much explosion energy can be released; it may depend on the rotation
of the BH and the progenitor.  The case of V4641 Sgr suggests the BH-
forming supernova was really {\sl dark}, because no radioactive
$^{56}$Ni was ejected.  Such a {\sl dark} supernova corresponds to the
extreme end of the {\sl faint} supernova branch \citep{nomoto05}.

Another possible scenario is the contamination by rotationally induced
mixing in the secondary star, itself. However, the observed rotational
velocity ($v \sin i \sim 100$~km s$^{-1}$) and estimated mass may be
lower than those predicted for the simultaneous enhancements of N and
Na, although they strongly depend on the uncertain inclination.

\vskip 5mm

We thank Drs. K. Matsumoto and M. Uemura for comments and suggestions.
This research was
partly supported by Grants-in-Aid from the Ministry of Education, Culture,
Sports, Science and Technology (No. 15540236 KS, No. 17540218 MTH, No.
17740163 NI, and No. 17033002 KN). T.C.B. ackowledges partial
support from grant AST 04-06784, as well as from grant PHY 02-16783, Physics
Frontier Center/Joint Institute for Nuclear Astrophysics (JINA), awarded by the
US National Science Foundation.

\twocolumn
\setcounter {table} {0}
\begin{table}
      \caption{Lines used in abundance analyses.}\label{first}
\footnotesize
      \begin{center}
      \begin{tabular}{cccc}
\hline\hline
          & Wavelength(\AA) & $\chi$(eV)  & log $\it gf$  \\
\hline
He~{\sc i} & 4471.469    & 20.964    & -2.198    \\
          &  4471.473    & 20.964    & -1.028    \\
          &  4471.473    & 20.964     & -0.278    \\
       &  4471.485    & 20.964    & -1.028    \\
          &  4471.488    & 20.964    & -0.548    \\
       &  4471.682    & 20.964    & -0.898    \\
        &  5875.599    & 20.964    & -1.511    \\
          &  5875.614    & 20.964    & -0.341    \\
          &  5875.615     & 20.964    &  0.409    \\
          &  5875.625    & 20.964    & -0.341    \\
          &  5875.640      & 20.964    &  0.139    \\
          &  5875.966    & 20.964    & -0.211    \\
C~{\sc ii} & 4267.001    & 18.046    &  0.562    \\   
       &  4267.261    & 18.046    & -0.584    \\
       &  4267.261    & 18.046    &  0.717    \\   
          &  4267.261    & 18.046    & -0.584    \\
N~{\sc i} &  6480.512    & 11.753    & -1.420    \\
       &  6481.706    & 11.750    & -1.063    \\
          &  6482.699    & 11.764    & -0.510    \\
          &  6483.753    & 11.753    & -0.857    \\
          &  6484.808    & 11.758    & -0.674    \\
O~{\sc i} &  6453.602    & 10.740    & -1.288    \\
          &  6454.444    & 10.740      & -1.066    \\
          &  6455.977    & 10.741     & -0.920    \\   
Na~{\sc i} & 5889.951    &  0.00     &  0.117    \\
       &  5895.924    &  0.00    &  -0.184   \\
Mg~{\sc i} & 5167.321    &  2.709    & -1.030    \\
          &  5172.684    &  2.712      & -0.402    \\
          &  5183.604    &  2.717    &  -0.180   \\
Mg~{\sc ii} & 4390.514    &  9.99    &  -1.490   \\
          &  4390.572    &  9.99       &  -0.530   \\
          &  4481.126    &  8.864    &   0.740   \\
          &  4481.325    &  8.864    &   0.590   \\
Al~{\sc ii} & 4663.046     &  10.598    &  -0.284   \\
Si~{\sc ii} & 5041.024    &  10.067    &   0.291   \\
          &  5055.984    &  10.074    &   0.593   \\
          &  5056.317    &  10.074    &  -0.359   \\
Ti~{\sc ii} & 4395.000    &   1.084    &  -0.510   \\
          &  4501.270    &   1.116    &  -0.760   \\   
          &  4571.960    &   1.572    &  -0.230   \\
Cr~{\sc ii} & 4558.660    &   4.073    &  -0.449   \\
          &  4588.220    &   4.071    &  -0.627   \\
          &  4592.070    &   4.074    &  -1.221   \\
Fe~{\sc ii} & 4508.280    & 2.856    & -2.250     \\
        &  4515.340    & 2.844    & -2.450     \\
          &  4576.330    & 2.844    & -2.920     \\
       &  4582.840    & 2.844    & -3.090     \\
       &  4583.830    & 2.807    & -1.860     \\
       &  4629.340    & 2.807    & -2.330     \\   
\hline
        \end{tabular}
      \end{center}
\end{table}
\setcounter {table} {1}
\begin{table}
      \caption{Abundances in V4641 Sgr.}\label{second}
      \begin{center}
      \begin{tabular}{cccccccccccc}
\hline\hline
  & C~{\sc ii}  & N~{\sc i}  & O~{\sc i} & Na~{\sc i} & Mg~{\sc i}  & Mg~{\sc ii} & Al~{\sc ii} &
Si~{\sc ii} & Ti~{\sc ii} & Cr~{\sc ii} & Fe~{\sc ii}  \\
\hline
[X/H] &  0.0 &    +0.8  &  0.0  & +0.8 & 0.0  &   +0.1    & +0.2  & 0.0
       &  0.0  &  0.0  & 0.0         \\
Error(A) & 0.12 &   0.02 &  0.01 & 0.12  & 0.13 &   0.03    & 0.04 & 0.03 &  0.10 &  0.05 & 0.04        \\
Error(B) & 0.27 &   0.0 &  0.01 & 0.16  & 0.18 &   0.19    & 0.14 & 0.06 &  0.03 &  0.10 & 0.13        \\
Error(C) & 0.01 &  0.02 &  0.0  & 0.13  & 0.10 &   0.02    & 0.08 & 0.05 &  0.13 &  0.09 & 0.10        \\
\hline
        \end{tabular}
      \end{center}
Note. --- [X/H]: logarithmic abundances relative to the Sun; Error (A), (B), and
(C): expected uncertainties in the abundance introduced by errors in
{\it $T_{\rm eff}$} (200 K), log {\it g} (0.5), and in $\xi$${}_{t}$ (0.5 km s${}^{-1}$),
respectively.        
\end{table}
\begin{figure}
      \begin{center}
        \FigureFile(80mm,85mm){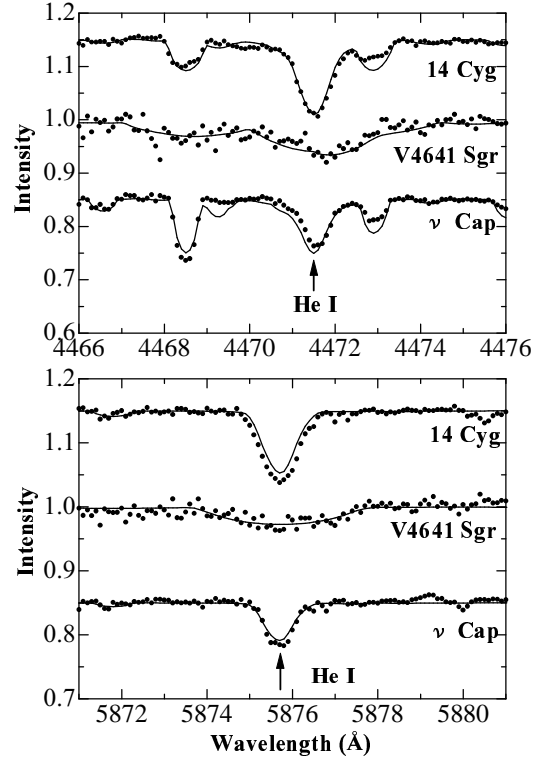}
      \end{center}
      \caption{Comparisons of He~{\sc i} lines. Profiles of two He~{\sc i} lines at
       4471 \AA~ and at 5876 \AA~ in V4641 Sgr and in two comparison
       stars 14 Cyg and in $\nu$ Cap are compared in the upper and 
       lower panels, respectively. The dots and solid lines
      show the observation and the simulated spectra calculated assuming the
      solar abundances, respectively. Rotational velocities, $\it v$ sin $\it i$, of
      30, 25, and 95 km s${}^{-1}$ are used for 14 Cyg, $\nu$ Cap, and V 4641 Sgr,
      respectively. }
     \end{figure}
\begin{figure}
      \begin{center}
        \FigureFile(80mm,65mm){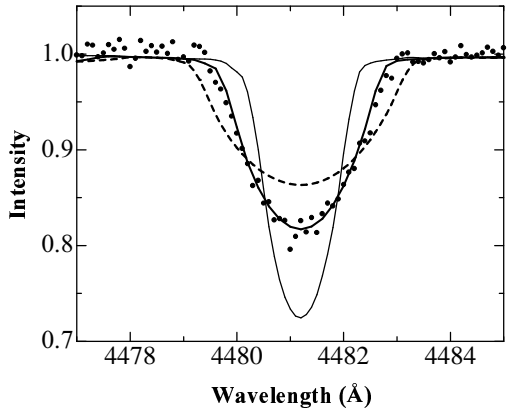}
      \end{center}
      \caption{Rotational velocity, $\it v$ sin $\it i$, of V4641 Sgr.
       A small section of the observed spectrum near the Mg~{\sc ii} line at
       4481.2 \AA~ is shown by dots. Simulated spectra,
       calculated assuming the abundance of Mg, [Mg/H] = 0.2, for $\it v$ sin $\it i$ =
       130, 95, and 60 km s${}^{-1}$ are shown by dashed, thick solid, and thin solid lines,
       respectively.}
     \end{figure}
\begin{figure}
      \begin{center}
        \FigureFile(80mm,75mm){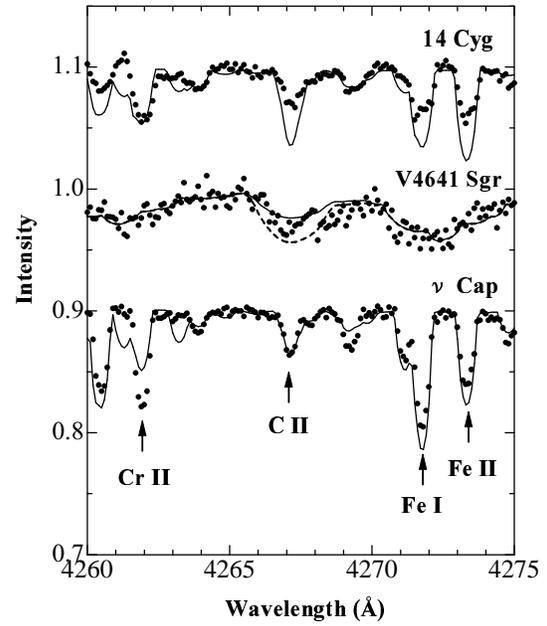}
      \end{center}
\caption{Analysis of the C~{\sc ii} feature at 4267.2 \AA. The  dots and thin solid lines
      show the observation and the simulated spectra calculated assuming the
      solar abundances of C. The thick  dashed line (for V4641 Sgr)
      show the case when C is over--abundant by 0.7 dex.}
     \end{figure}

\begin{figure}
      \begin{center}
        \FigureFile(80mm,75mm){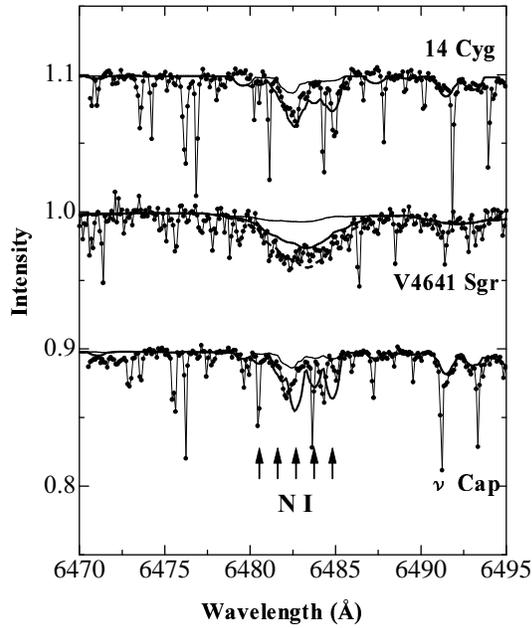}
      \end{center}
\caption{Analysis of the N~{\sc i} feature near 6483 \AA. The dots and thin solid lines
      show the observation and the simulated spectra calculated assuming the
      solar abundances of N. Many sharp absorption features are caused by
      atmospheric H${}_{2}$O molecules. The thick solid lines  and the dashed line (for V4641 Sgr)
      show the case when N is over--abundant by 1.0 dex and 1.5 dex, respectively.}
     \end{figure}
\begin{figure}
      \begin{center}
        \FigureFile(80mm,75mm){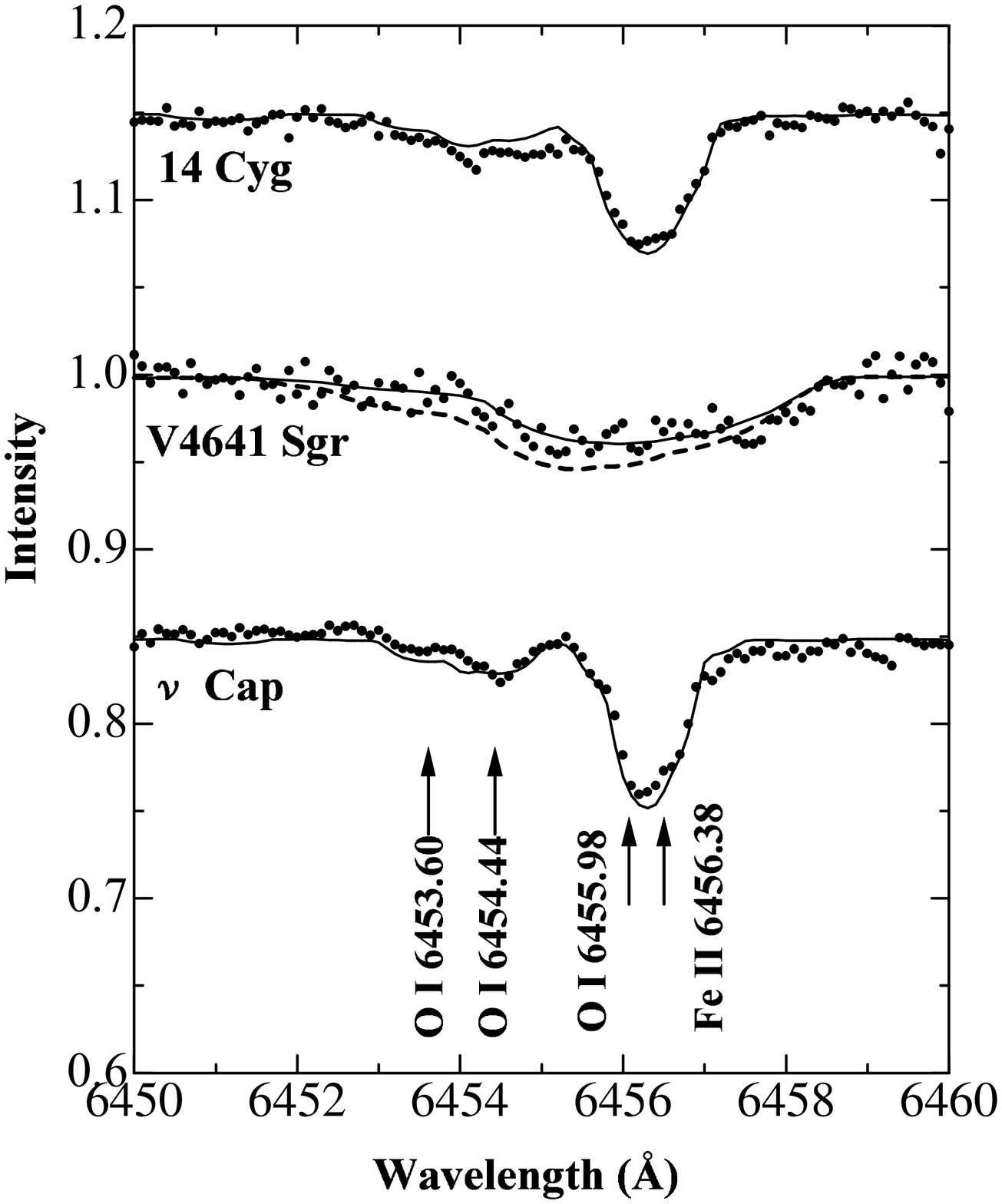}
      \end{center}
      \caption{Analysis of O~{\sc i} lines near 6455 \AA. The dots and solid lines
      show the observation and the simulated spectra calculated assuming 
      solar abundances of O and Fe, respectively. The dashed line for V4641 Sgr is the
      case when O is 0.5 dex over--abundant as noted in \citet{orosz01}.}
     \end{figure}

\begin{figure}
      \begin{center}
        \FigureFile(80mm,75mm){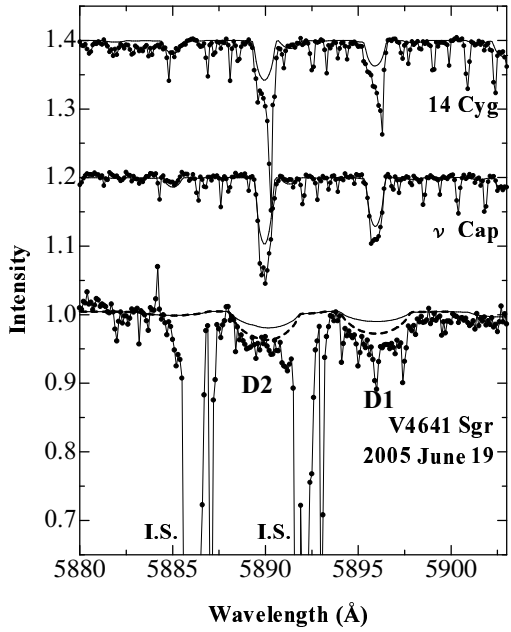}
      \end{center}
\caption{Analysis of the Na~{\sc i} D lines. Both D1 and D2 lines in 14 Cyg and
         $\nu$ Cap are contaminated by sharp interstellar (I.S.) components.
          The I.S. lines of V4641 Sgr observed on 2005  June 19 are apparently
          blue-shifted  by 210 km s${}^{-1}$ due to the orbital motion of the star.
          The dots and thin solid lines
         show the observation and the simulated spectra calculated assuming the
         solar abundances of Na. The dashed line for V4641 Sgr shows the case when
         Na is over--abundant by 0.7 dex.}
     \end{figure}
\begin{figure}
      \begin{center}
        \FigureFile(80mm,75mm){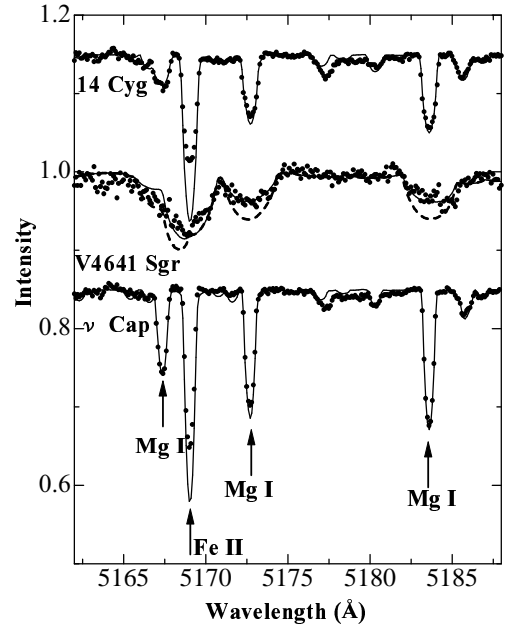}
      \end{center}
      \caption{Analysis of the Mg~{\sc i} triplet lines. Dots and solid lines
      show the observation and the simulated spectra calculated assuming 
      solar abundances of Mg and Fe, respectively. The dashed line for V4641 Sgr is the
      case when Mg is 0.85 dex over--abundant as noted in \citet{orosz01}.}
     \end{figure}

\begin{figure}
      \begin{center}
        \FigureFile(80mm,75mm){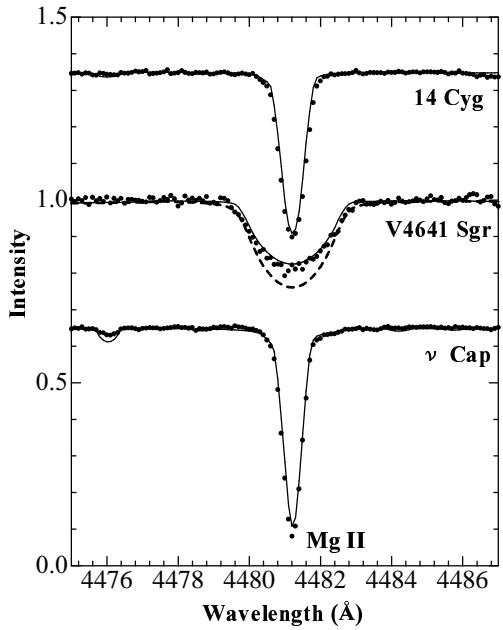}
      \end{center}
      \caption{Analysis of the Mg~{\sc ii} line at 4481 \AA. The dots
       represent observed spectra. The solid lines for 14 Cyg and  $\nu$ Cap are
      simulated spectra calculated assuming the  solar abundance of Mg,
      while that for V4641 Sgr is calculated assuming an enhanced (by 0.20 dex)
      abundance of Mg. The dashed line for V4641 Sgr is the
      case when Mg is 0.85 dex over--abundant as noted in \citet{orosz01}.}
     \end{figure}

\begin{figure}
      \begin{center}
        \FigureFile(80mm,75mm){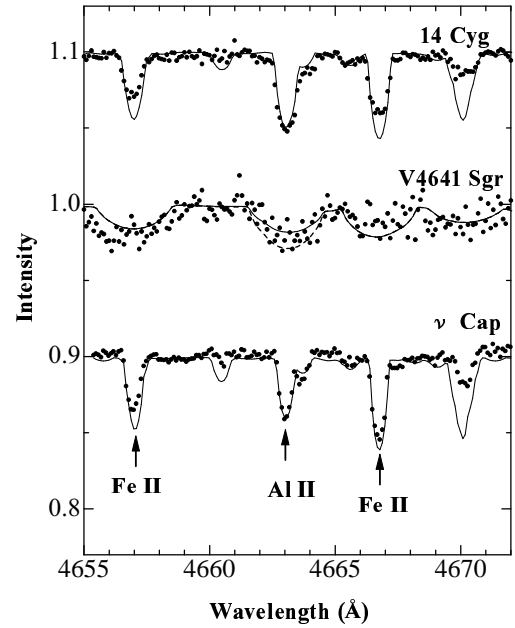}
      \end{center}
\caption{Analysis of the Al~{\sc ii} line at 4663 \AA. The dots and solid lines
      show the observation and the simulated spectra calculated assuming a
      solar abundances for Al, respectively. The dashed line for V4641 Sgr is the
      case when Al is 0.5 dex over--abundant.}
     \end{figure}
\begin{figure}
      \begin{center}
        \FigureFile(80mm,75mm){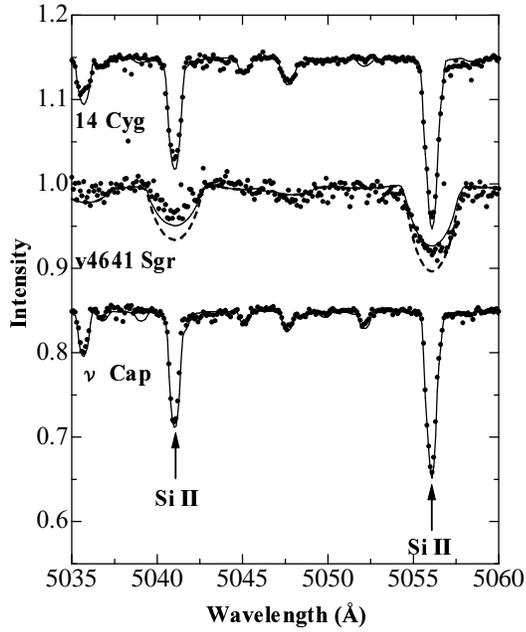}
      \end{center}
      \caption{Analysis of a pair of Si~{\sc ii} lines near 5050 \AA. The dots and solid lines
      show the observation and the simulated spectra calculated assuming the
      solar abundance of Si, respectively. The dashed line for V4641 Sgr is the
      case when Si is 0.5 dex over--abundant.}
     \end{figure}
\begin{figure}
      \begin{center}
        \FigureFile(80mm,75mm){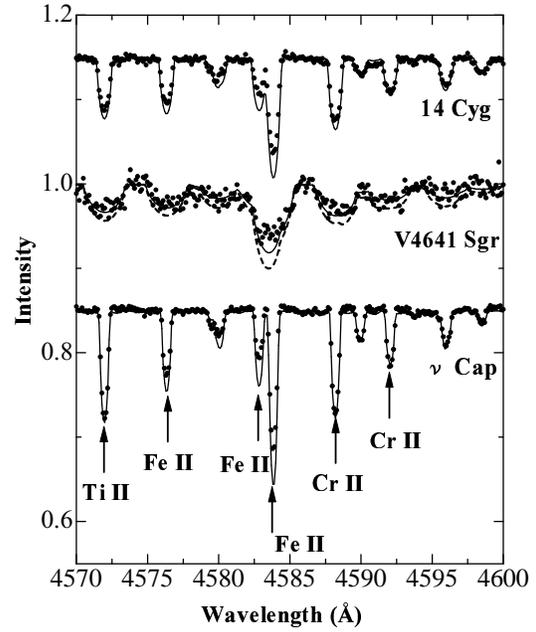}
      \end{center}
      \caption{Analysis of Ti~{\sc ii}, Cr~{\sc ii}, and Fe~{\sc ii} lines near
      4585 \AA. The dots and solid lines
      show the observation and the simulated spectra calculated assuming 
      solar abundances of Ti, Cr, and Fe, respectively. The dashed line for V4641 Sgr is the
      case when Ti, Cr, and Fe are  0.2 dex over--abundant.}
     \end{figure}
\begin{figure}
      \begin{center}
        \FigureFile(80mm,75mm){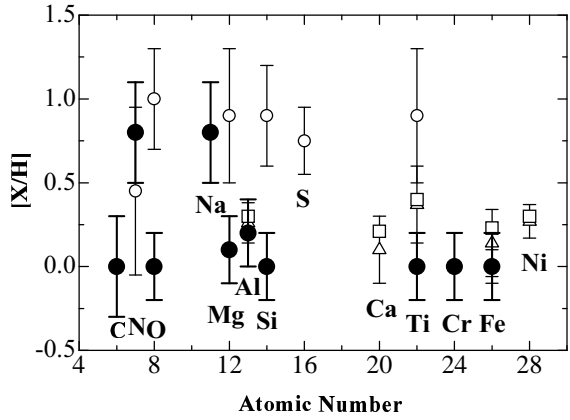}
      \end{center}
      \caption{Comparison of metal abundances. [X/H] values, logarithmic
      abundance of the element X relative to the Sun, in four
      secondary stars are plotted against the atomic number. Filled
      circles, open circles, open triangles, and open squares are
      for V4641 Sgr, GRO J1655-40, A0620-00, and for Cen X-4, respectively. }
     \end{figure}

\begin{figure}
      \begin{center}
        \FigureFile(80mm,80mm){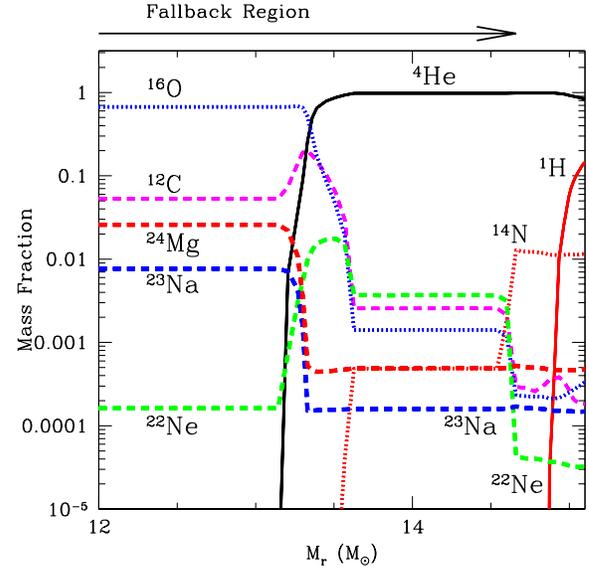}
      \end{center}
      \caption{Internal abundance distribution of the solar metallicity
 15.14$M_\odot$ He star model at the onset of collapse.  This He star is the core
 of the 40$M_\odot$ star.  The extent
 of the large scale fallback is $M_r=14.66M_\odot$ shown by
 right-headed arrow. The remaining black hole mass is $14.66M_\odot$. The main
 contribution for Na as well as for N comes from H-burning. N and Na
 abundances in the He layers are enhanced with respect to the
 initial abundances of $1.2 \times 10^{-3}$ and $3.5 \times 10^{-5}$,
 respectively.  }
     \end{figure}

\begin{figure}
      \begin{center}
        \FigureFile(80mm,80mm){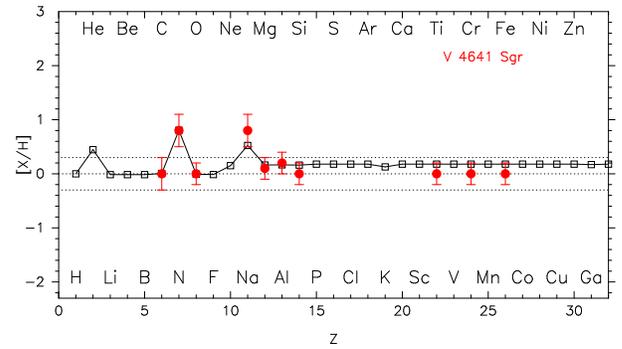}
      \end{center}
      \caption{Comparison of V4641~Sgr with a mixture of the solar
 metallicity $15.14M_\odot$ He star supernova ejecta and the secondary
 star materials. The assumed ratio of the supernova ejecta to the
 secondary star materials is 1/40. The filled circles with error-bars show
 the observed abundances. The abundance pattern of the mixture is
 represented by open squares connected with solid lines. The yield
 shows that, except for the enhancements of N and Na, all the other
 elements have almost solar ratios. Black hole with the mass of
 $14.66M_\odot$ is left behind. }
     \end{figure}


\begin{thebibliography}{}
\bibitem[Abt et al.(2002)]{abt95a}
         Abt, H. A., Levato, H., \& Grosso, M. \  2002, \apj,  573, 359

\bibitem[Abt and Morrell(1995)] {abt95b}
         Abt, H. A., \& Morrell, N. I. \ 1995,
         \apjs,  99, 135

\bibitem[Adelman(1991)] {adel91}
         Adelman, S. J. \ 1991, \mnras, 252, 116

\bibitem[Adelman(1999)] {adel99}
         Adelman, S. J. \ 1999, \mnras, 310, 146

\bibitem[Adelman et al.(2002)] {adel02}
         Adelman, S. J., Pintado, O. I., Nieva, F., Rayle, K. E., \&
        Sanders, S. E., Jr.  \ 2002, \aap, 392, 1031

\bibitem[Dworetsky and Budaj(2000)]{dwo00}
        Dworetsky, M. M., \& Budaj, J. \ 2000, \mnras, 318, 1264

\bibitem[Fekel(2003)]{fekel03}
        Fekel, F. C. \ 2003, \pasp, 115, 807


\bibitem[Gonz\'alez Hern\'andez et al.(2004)]{gonz04}
         Gonz\'alez Hern\'andez, J. I., Rebolo, R., Israelian G., \& Casares, J.
         \ 2004, \apj, 609, 988

\bibitem[Gonz\'alez Hern\'andez et al.(2005)]{gonz05}
         Gonz\'alez Hern\'andez, J. I., Rebolo, R., Israelian G.,  Casares, J.,
         Maeda, K., Bonifacio, P., \& Molaro, P.
         \ 2005, \apj, 630, 495

\bibitem[Goranskij et al.(2003)]{goran03}
         Goranskij, V. P., Barsukova, E. A., \& Burenkov, A. N. \ 2003,
         Astronomy Rept., 47, 740

\bibitem[Grevesse and Sauval(1998)]{gre98}
        Grevesse, N., \& Sauval, A. J. \ 1998, Space Sci. Rev., 85, 161

\bibitem[Israelian et al.(1999)]{isra99}
         Israelian, G., Rebolo, R., Basri, G., Casares, J., \&
         Martin, E. L. \ 1999, Nature, 401, 142

\bibitem[Iwamoto et al.(2005)]{iwamoto05}
        Iwamoto, N., Umeda, H., Tominaga, N., Nomoto, K., \& Maeda, K.
        \ 2005, Science, 309, 451



\bibitem[Kato et al.(1999)]{kato99}
         Kato, T., Uemura, M., Stubbings, R., Watanabe, T., \&
         Monard, B. \ 1999, Inf. Bull. Variable Stars, 4777

\bibitem[Kupka et al.(1999)]{kupka}
         Kupka, F., Piskunov, N., Ryabchikova, T. A., Stempels, H.C.,
         \& Weiss, W.W. \ 1999, \aaps, 138, 119


\bibitem[Kurucz(1993)]{kuru93}
         Kurucz, R. L. \ 1993, Kurucz CD-ROM, No.13 (Harvard-Smithsonian Center
         for Astrophysics)

\bibitem[Kurucz and Bell(1995)] {kubell95}
         Kurucz, R. L., \& Bell, B. \ 1995, Kurucz CD-ROM, No.23
        (Harvard-Smithsonian Center  for Astrophysics)



\bibitem[Noguchi et al.(2002)]{nogu02}
         Noguchi, K., et al.  \ 2002,   \pasj, 54, 855

\bibitem[Nomoto et al.(2005)]{nomoto05}
         Nomoto, K., Tominaga, N., Umeda, H., Maeda, K., Ohkubo, T., \& Deng,
         J. \ 2005, Nuclear Phys., A 758, 263c



\bibitem[Orosz et al. (2001)]{orosz01}
         Orosz, J. A., et al. \ 2001, \apj, 555, 489 

\bibitem[Podsiadlowski et al. (2002)]{podsi}
         Podsiadlowski, P., Nomoto, K, Maeda, K., Nakamura, T., Mazzali, P.,
         \& Schmidt, B. \ 2002, \apj, 567, 491

\bibitem[Royer et al.(2002)]{royer02}
         Royer, F., Grenier. S., Baylac, M. -O., Gomez, A. E., \&
         Zorec, J.  \ 2002, \aap, 393, 897

\bibitem[Smith et al. (1999)]{smith99}
         Smith, D. A., Levine, A. M., \& Morgan, E. H.,
        \ 1999, IAU Circ., 7253, 2

\bibitem[Takada-Hidai, Aoki, and Zhao (2002)]{hidai02}
        Takada-Hidai, M., Aoki, W., \& Zhao, G. \ 2002, \pasj, 54, 899

\bibitem[Uemura et al.(2002a)]{uemura02a}
         Uemura, M., et al. \ 2002a, \pasj, 54, 95

\bibitem[Uemura et al.(2002b)]{uemura02b}
         Uemura, M., et al. \ 2002b, \pasj, 54, L79

\bibitem[Uemura et al.(2004a)]{uemura04a}
         Uemura, M., et al. \ 2004a, \pasj, 56, S61

\bibitem[Uemura et al.(2004b)]{uemura04b}
         Uemura, M., et al. \ 2004b, \pasj, 56, 823

\bibitem[Uemura et al.(2005)]{uemura05}
         Uemura, M., et al. \ 2005, Inf. Bull. Variable Stars, 5626

\bibitem[Umeda and Nomoto (2003)]{umeda03}
         Umeda, H., \& Nomoto, K. \ 2003, Nature, 422, 871

\bibitem[Umeda and Nomoto (2005)]{umeda05}
         Umeda, H., \& Nomoto, K. \ 2005, \apj, 619, 427


\end{thebibliography}
\end{document}